\documentclass{article}
\usepackage{PRIMEarxiv}

\pdfoutput=1 

\usepackage{amsmath} % assumes amsmath package installed
\usepackage{amssymb}  % assumes amsmath package installed
\usepackage[utf8]{inputenc} % allow utf-8 input
\usepackage[T1]{fontenc}    % use 8-bit T1 fonts
\usepackage{hyperref}       % hyperlinks
\usepackage{url}            % simple URL typesetting
\usepackage{booktabs}       % professional-quality tables
\usepackage{amsfonts}       % blackboard math symbols
\usepackage{nicefrac}       % compact symbols for 1/2, etc.
\usepackage{microtype}      % microtypography
\usepackage{fancyhdr}       % header
\usepackage{graphicx}       % graphics
\graphicspath{{media/}}     % organize your images and other figures under media/ folder

\usepackage{enumitem}

\newcommand{\cost}{{\tt cost}}
\newcommand{\loss}{{\tt loss}}

\title{Designing Fair, Cost-optimal Auctions based on Deep Learning for
Procuring Agricultural Inputs through Farmer Collectives
}

\author{
  Mayank Ratan Bhardwaj, Bazil Ahmed and Prathik Diwakar \\
  Indian Institute of Science \\
  Bengaluru\\
  \texttt{\{mayankb, bazilahmed, prathikd\}@iisc.ac.in} \\
   \And
  Ganesh Ghalme  \\
  Indian Institute of Technology \\
  Hyderabad\\
  \texttt{ganeshghalme@ai.iith.ac.in} \\
  \And
  Y. Narahari  \\
  Indian Institute of Science \\
  Bengaluru\\
  \texttt{narahari@iisc.ac.in} \\
}

\begin{document}
\maketitle

\begin{abstract}
Procuring agricultural inputs (agri-inputs for short) such as seeds, fertilizers, and pesticides, at desired quality levels and at affordable cost, forms a critical component of agricultural input operations. This is a particularly challenging problem being faced by small and marginal farmers in any emerging economy. Farmer collectives (FCs), which are cooperative societies of farmers, offer an excellent prospect for enabling cost-effective procurement of inputs with assured quality to the farmers. In this paper, our objective is to design sound, explainable mechanisms by which an FC will be able to procure agri-inputs in bulk and distribute the inputs procured to the individual farmers who are members of the FC. In the methodology proposed here, an FC engages qualified suppliers in a competitive, volume discount procurement auction in which the suppliers specify price discounts based on volumes supplied. The desiderata of properties for such an auction include: minimization of the total cost of procurement; incentive compatibility; individual rationality; fairness; and other business constraints. An auction satisfying all these properties is analytically infeasible and a key contribution of this paper is to develop a deep learning based approach to design such an auction. We use two realistic, stylized case studies from chili seeds procurement and a popular pesticide procurement to demonstrate the efficacy of these auctions.
\end{abstract}


\section{Introduction}
Sourcing the right quality and quantity of agricultural inputs (agri-inputs for short) such as seeds, fertilizers, pesticides, farm equipment, and human resources is a  critical aspect of agricultural operations. This is a universal problem faced by farmers throughout the globe and especially in emerging economies. 
\newline \newline 
\noindent
{\bf Agri-Inputs:} In order to maximize yield, it is necessary for farmers to use the right  quantities of right quality inputs. 
Agricultural inputs are external inputs required to help the various farming operations. They range from high-quality seeds, fertilizers, pesticides, to high-tech tractors and farm equipment.
There are mainly two categories of  agri-inputs:  consumable and capital. Consumable inputs are used commonly and regularly by the  farmers - seeds, fertilizers, pesticides, etc. These are essentially natural materials that will be consumed by the crops. 
Capital inputs include farm equipment such as tractors, agricultural robots, trellising materials, etc. %and other gardening infrastructure. 
In addition to all these, we have human labor.  
In this paper, we focus on consumable inputs.

Seeds that are usually procured include: vegetables, cotton, paddy, maize, sorghum, sunflower, wheat, millets, mustard, chili peppers, and lentils. 
High quality seeds facilitate smooth farming; low quality seeds can lead to crop losses and even crop destruction. 
Pesticides used by farmers belong to the following categories: (a) herbicides, %(to control weeds and other plants),  
(b) insecticides, %(to control insects),  
(c) fungicides, %(to control fungi or other plant pathogens), 
(d) nematicides, %(to control parasitic worms), 
and (e) rodenticides. %(to control rodents). 
The use of pesticides in right quantities at the right time saves the crops from being destroyed. Fertilizers are any materials of natural or synthetic origin that are applied to soil or to plant tissues to supply plant nutrients. 
%Prominent ones are based on Nitrogen, Phosphorous, Potassium, Calcium, Magnesium, and Sulphur.  \\[1mm]
\newline\newline
\noindent
{\bf The Context and Need:} In many emerging economies, most of the farmers are small or marginal, holding less than 5 acres of land. Their economic condition is weak and they mostly depend on credit for sustaining their operations. 
The farmers are faced with a dilemma at the start of every cropping season over where to procure agri-inputs from. Most of these farmers, being small, buy the inputs on credit. It is extremely important for them to get quality inputs, but, often, they end up with low quality inputs  leading to crop losses or even crop failures. To quote a typical farmer in one of the emerging economies:
``If inputs are substandard or fake, we have to go through a minimum of two years of suffering due to debt. Therefore, in addition to the good rains, the quality of inputs we buy decides our farm income season after season." In fact, approximately 50 percent of the total cost goes towards inputs \cite{HAPPENINGS20}.

A seemingly simple and promising way in which this problem could be solved is to create economies of scale in procuring high quality inputs  through a  collective action by setting up a bulk procurement system. 
Bulk procurement also makes it easier to ensure that the quality of input is maintained. In many countries, the respective governments have taken up the initiative to launch farmer collectives or farmer cooperatives (FCs) to help out the small and marginal farmers in various input and output operations. In particular, FCs would be extremely helpful for reducing  the input burden on the farmer through bulk procurement of inputs, after collecting information on the input requirements of individual farmers (this could be done, for example, by a well designed mobile app to reach out to all farmer members of the FC).  This paper specifically focuses on this problem and explores the use of systematic methods for harnessing the bargaining  power of farmer collectives. 
%Collective purchase of agri-inputs followed by supplying them at affordable prices to the farmers reduces the per unit cost while ensuring quality of the inputs. 
%Collectives can also provide extension services such as cleaning, packaging, quality assurance, credit facility, etc.  Such farmer collectives have the additional benefit of assuring better prices for the produce through collective sale of agricultural produce to a bigger market.  

There exist  more than 1500 FCs in Brazil, more than 7500 FCs in Germany, more than 63000 FCs in India, and more than 1700 FCs in USA. %To gauge the quantum of impact such collective action for agri-input procurement could possibly have, our survey on the number of farmer collectives in different countries reveals, for example, more than 1500 FCs in Brazil, more than 7500 FCs in Germany, more than 63000 FCs in India, and more than 1700 FCs in USA. 
A recent report \cite{EarthObservingSystem22} mentions a staggering 1.2 million agricultural cooperatives across the globe today. It may not be unreasonable to expect a saving of US \$ 100 for procuring a typical seed  variety in a typical FC which may consist of a few hundreds of farmers. Thus, such collective action for agri-input procurement has the potential for huge impact.

To gain a first-hand experience and knowledge of the real situation on the ground, our research group undertook a field study of two FCs, Anekal Horticulture Producers Company Limited and Rajaghatta Horticulture Farmer Producer Company Limited, both within 50 km from Bengaluru, India. Both FCs have a membership of about 1000 farmers each.
Small and marginal farmers tend to be low on education and are particularly vulnerable to the strategic tactics of intermediaries. Since the intermediaries offer credit to the farmers for sourcing the inputs, the intermediaries are able to wield their influence in the marketing and selling of the produce as well. In the process,
the farmers end up on the losing side. The FCs play a key role in streamlining the supply of inputs to the farmers and can counter the selfish moves of the intermediaries.
Our team had a detailed discussion with the FCs on how they collect indents, aggregate the input requirements of the farmers and bulk-procure the right quantities of inputs to be sold subsequently to the farmers at affordable prices.
During these conversations, we also realized numerous issues which were hampering a successful execution of the bulk
procurement process. For example, FCs do need a healthy amount of money and resources to execute this process more
efficiently. We also discussed with the FCs the discounts that they would be able to obtain because of both volume and
variety of their purchases. Here again, we found that the discounts on offer from the suppliers could be much higher if
the right kind of procurement mechanisms are put in place.

Note that there are numerous types and varieties of seeds. There are numerous types of pesticides and fertilizers. The total savings, by conservative estimates, easily runs into Billions of Dollars (a more methodical and accurate estimate of this is beyond the scope of this paper). Above all, the farmers, who are often naive, are insulated from the transactions and further, the quality levels of agri-inputs are assured.  

Since scientifically designed auction mechanisms can promote honest behavior and healthy competition among suppliers \cite{MILGROM89}, we propose to develop a suitable procurement auction (also called reverse auction) mechanism 
for bulk procurement of agri-inputs by FCs. 
%In this paper, our main objective is to develop a suitable procurement auction mechanism \cite{HOHNER03, BICHLER05, CHANDRASHEKAR07, IYENGAR08, GAUTAM09, BHARDWAJ22} for bulk procurement of agri-inputs by FCs. \newline \newline
%Many State Governments in India have schemes for preferential procurement of produce from farmer collectives.
\newline \newline
\noindent
{\bf Procurement Auctions for Bulk Procurement by Farmer Collectives:}  We explain the workflow of the proposed agri-input procurement with an example. Let us consider procurement of paddy seeds. Suppose the farmer collective has 1000 farmer members (like Anekal or Rajghatta described above) who want to grow paddy and suppose there are five varieties of paddy seeds. The FC will collect from the farmers (this can be done using a mobile app), the quantum and type of seeds required by each farmer. The FC will aggregate these requirements into five buckets, each bucket corresponding to a particular seed variety. Each bucket will then correspond to a certain large volume of seeds of that variety needed by all the farmers. For each bucket, the FC will identify suppliers for the seeds, qualify them based on the reputation and quality standards,  and conduct an auction involving the qualified suppliers.

First, we bring out the most desirable properties for such an auction mechanism. Our analysis is based on our familiarity with and study of FCs and agri-input procurement. 
\begin{enumerate}[leftmargin = *]
\item  {\bf Incentive compatibility  (IC)} \\Incentive compatibility ensures truthful bidding by the suppliers and is a fundamental requirement of any auction mechanism. The most powerful version of IC is dominant  strategy incentive compatibility (DSIC), which means bidding true values is best irrespective of the bids of the other players.
\item 
{\bf Individual rationality (IR)}  \\Individual rationality ensures that the suppliers obtain non-negative utility by participating in the auction. The most powerful version of IR is ex-post IR, which implies that the utility to each participating player will be non-negative irrespective of the bids of the other players.
\item {\bf Social welfare maximization (SWM)}  \\This implies maximizing the sum of utilities of the participants in the auction. In a procurement auction, the participants involved are the FC and the suppliers. 
\item {\bf Cost minimization (OPT)} \\This means the expected total cost of procurement is minimized by the auction. Cost minimization is a key requirement since we wish to minimize the input costs to the small and marginal farmers. 
\item {\bf Fairness (FAIR)}   \\Fairness implies that the winning suppliers are chosen in a fair way. An index of fairness would be envy-freeness -- no supplier can increase her utility by adopting another supplier's outcome. If envy-freeness is not achievable, the next best option is envy minimization (which is what we pursue in this paper).
\item {\bf Business constraints (BUS)}   \\Satisfaction of business constraints refers to constraints such as having a minimum number of winning suppliers (to avoid monopoly),  a maximum number of winning suppliers (to minimize logistics costs), a maximum fraction of business to be awarded to any supplier, etc. 
\end{enumerate}

Further, the auction should extract volume discounts or quantity discounts from suppliers. This final requirement points to using auctions where the suppliers can specify volume discounts as a part of their bids.

Volume discount auctions or quantity discount auctions,
which are very relevant to this work are covered in \cite{HOHNER03, BICHLER05, CHANDRASHEKAR07, IYENGAR08, GAUTAM09}. In a volume discount auction, the buyer wishes to procure a large volume of  a certain item type and the suppliers announce {\em supply curves\/} or {\em volume discount bids\/}.  A supply curve specifies the discounted prices offered by the supplier. \\[1mm]

\noindent
{\bf Example 1}: 
Suppose we are procuring paddy seeds in packets (usually a packet may correspond to a certain number of kilograms). Let an  example supply curve be ((1-500: 20), (501-1500: 18), (1501-2500: 16)). This would mean: the supplier offers a per unit price of \$20  when the quantity is in the range 1-500; a per unit price of \$18  when the quantity is in the range 501-1500; and a per unit price of \$16 when the quantity is in the range 1501-2500. If the seller is supplying 1800 units, the total bid price will be $500 \times \$20 + 1000 \times \$18 + 300 \times \$16 = \$32800$. Our field studies have shown that such supply curves are quite common in bulk procurement of agri-inputs. In this paper, we are concerned with design of such auctions.\newline

\noindent
{\bf Contributions and Outline:} A recent paper \cite{BHARDWAJ22} has addressed the problem of procuring agri-inputs through farmer collectives by proposing volume discount auctions and combinatorial auctions. \cite{BHARDWAJ22}  proposes  Vickrey-Clarke-Groves (VCG) payments to ensure that the auction is dominant strategy incentive compatible and maximizes social welfare. Their auction also takes into account certain business constraints such as minimum number of winning suppliers and maximum number of winning suppliers. Their auction is based on the methods described in \cite{HOHNER03, BICHLER05, CHANDRASHEKAR07, IYENGAR08, GAUTAM09} and performs better than the naive methods which are currently being used.  However, their auction crucially does not minimize the total cost of procurement and does not implement any notion of fairness. 

Satisfying all of the six properties listed above (IC, IR, SWM, OPT, FAIR, BUS) is clearly a tall order. Mechanism design theory \cite{KRISHNA09, NARAHARI14} is replete with impossibility theorems which make it clear that these properties  cannot be simultaneously satisfied. The ambitious goal of this paper is to innovatively devise auctions that achieve maximal subsets of these objectives  with as little compromise as possible. 

%, a crucial requirement for procuring agri-inputs, especially with small and marginal farmers involved. Also, in the agriculture and other contexts, it would be important to satisfy certain fairness constraints (such as envy-freeness of allocations, fair allocation of business to competing suppliers, etc.). 
%Thus the desirable desiderata for an agri-input auction include: (a) incentive compatibility (b) individual rationality (c) social welfare maximization (d) cost minimization (e) fairness  (f) satisfaction of  certain business constraints. Section 2 describes these properties  formally.
%\textcolor{red}{Should we refer to section 2 for formal definitions?}

The methodology proposed in this paper hinges on our technically novel extension to a recent line of research \cite{FENG18, DUTTING21, ZHANG21} that explores a deep learning approach to optimal auction design and achieves, in the sense of regret minimization, properties IC, IR, and OPT. Our methodology considers volume discount bids and achieves, properties FAIR and BUS in addition, with a slight compromise in  SWM. %SWM, which is less important than OPT for our problem setting.
This is a key technical contribution of this paper, which is described in the next section (Section \ref{sec:DLApproach}).
%It is notable that the RegretNet architecture methodology presented in  \cite{DUTTING21} cannot handle non-additive valuations which are characteristic of volume discount auctions. There are a couple of issues because of which RegretNet cannot be directly used for our context. Our methodology takes care of these issues and a key technical contribution of our paper. We present the details of this methodology in  Section \ref{sec:DLApproach} of the Paper.

%Using this methodology, we show how to implement a sequence of increasingly powerful volume discount auctions: (a) Cost minimization  (b) Cost minimization with envy minimization (c) Cost minimization with business constraints (d) Cost minimization with envy minimization as well as business constraints. All these four auctions are subject to dominant strategy incentive compatibility (strongest notion of incentive compatibility) and ex-post individual rationality (strongest notion of individual rationality). 

The design of the auction mechanism above is clearly motivated by the agri-input procurement by farmer collectives. We demonstrate the efficacy of these mechanisms through thought experiments in Sections \ref{sec:ExperimentalResults} and \ref{sec:CaseStudies}.  In Section \ref{sec:ExperimentalResults}, we conduct experiments on data that are synthetically generated based on real-world observations of the volume discounts offered by different suppliers. 
%We present results for all the proposed auctions and present insights from the numerical results. In particular, we show that all these auctions outperform the corresponding VCG auctions.
In Section \ref{sec:CaseStudies}, we describe two realistic, stylized  case studies: (1) procurement of chili pepper seeds (2) procurement of a popular, commonly used pesticide. We describe our experiments on these case studies; the results again highlight the superior performance of the proposed mechanism. 
Finally, Section \ref{sec:Conclusion} concludes the paper and provides some directions for future work.
 
%\textcolor{red}{Should we be mentioning "in our vicinity" when considering that we mentioned the previous paper (Bhardwaj et al)? It might be considered deanonymizing} Just the fact that we are working on the same example (without even mentioning that the authors of that paper had shared the data publicly) might hint at our identity
% Prathik: True. Please see if you can figure out a way around this, as half of the experiments rely on this while the other half are based on novel data

\section{A Deep Learning Approach for Volume Discount Procurement Auctions}\label{sec:DLApproach} 
%\section{A DEEP LEARNING APPROACH FOR VOLUME DISCOUNT PROCUREMENT AUCTIONS}\label{sec:DLApproach} 
As stated already, a volume discount auction satisfying DSIC, IR, OPT, FAIR, and BUS is impossible in mechanism design \cite{KRISHNA09, NARAHARI14}. To surmount this, we adopt a deep learning approach, which solves this design problem as a regret minimization problem on the lines of \cite{DUTTING21}.  Even though  our architecture  resembles the RegretNet architecture proposed by \cite{DUTTING21}, it is notable that the RegretNet architecture cannot be applied as an off-the-shelf mechanism for volume discount procurement auctions.  There are two technical reasons for this. 
%In fact, it requires technically novel extensions.   %\sout{The RegretNet architecture in \cite{DUTTING21} cannot be applied directly to our setting which requires all the above properties to be satisfied. We first bring out the issues  involved in using the Regretnet architecture for our problem setting and next describe our methodology.}
%
%\paragraph{RegretNet cannot be used for volume discount auctions}\label{subsec:Challenges}

 The first reason has to do with satisfying individual rationality.  RegretNet is designed for forward auctions (selling scenario) while the procurement auction is a reverse auction. IR requires that all agents receive a higher utility by participating in the auction than by not participating. In a forward auction, this means that no buyer should be paying more than their valuation of the items allocated to them. In a procurement auction, on the other hand, this means that no seller would accept a payment lower than their valuation for the items they are selling. For the procurement auction setting, the network requires a subtle modification (see Section \ref{subsec:NetDetails})  to satisfy the IR condition.
 
The second reason is more fundamental. We are dealing with the procurement of homogeneous units with volume discounts whereas the work of \cite{DUTTING21}  considers auctions with additive (or unit-demand)  valuations.  Our volume discount auction setting is not additive and is not unit-demand either. To see this, observe that the value  of $2x$ units with volume discounts is not the same as twice the value of $x$ units. Additionally, the suppliers in our setting wish to sell any number of units, unlike the unit-demand buyers considered by \cite{DUTTING21}. Auctions with additive valuations and unit-supply valuations make it possible to use allocation networks whose outputs are simply (stochastic) allocation matrices. We simply cannot do this in our case, so instead, we produce an allocation tuple as output - with each element in the tuple being the allocation for the corresponding supplier. This complicates the computation of the payments and makes theoretical analysis significantly complex. 

\subsection{The Volume Discount Procurement Auction Setting}\label{subsec:Setting}
%\textcolor{red}{I have moved all references to Dutting's paper to a separate section now. We can also cite other papers there, if anyone has any ideas. Also, notation has been changed, please have a look}
% Prathik: Everyone please check this section of the paper and add a comment if everything is OK
%\textcolor{red}{PS: Check this section}
In this section, we formally describe the volume discount auction setting. There is a single buyer who intends to procure $m$ homogeneous units of a certain item from $n$ suppliers using a procurement auction with volume discount bids. Let $\ell := \lfloor \frac{m}{k} \rfloor$ for some predefined $k$. %and let $g_j$ denote the $j$th good according to arbitrary but fixed   indexing over the goods.
The volume discount bidding  is implemented as follows. First, each supplier $i$  submits a volume discount bid in the form of a vector $b^{(i)} = (b^{(i)}_1, b^{(i)}_{2}, \cdots, b^{(i)}_{k})$ of $k$ intervals (See Remarks subsection below for details).
%Should we explain what this(volume discount bid) means? Or at least mention that it is explained in Remarks? Should we make Remarks a numbered subsection?
% Prathik & Bazil: Yes, I think we should mention that we have explained volume discounts better in the remarks
Then, given the vector of bids $b = (b^{(1)}, \cdots , b^{(n)})$ as input, the procurement mechanism outputs allocation and payment vectors denoted by the tuple
% Prathik: Is angled brackets necessary here? We have used parentheses for tuples everywhere else. Ganesh?
 $\langle a(b), p(b) \rangle$. Here,  $a(b)   = (a_1(b), \cdots , a_n(b))$ denotes the allocation vector  and $p(b) = (p_1(b), \cdots , p_n(b))$ denotes the payment vector with each  $a_i(b)$ being the number of units bought from supplier $i$ and $p_i(b)$ being the payment made to supplier $i$. 
  Note that such auctions or variants have been considered earlier in \cite{HOHNER03, BICHLER05, CHANDRASHEKAR07, IYENGAR08, GAUTAM09}. In this paper we seek nontrivial extensions, to satisfy additional properties such as FAIR and BUS.

The suppliers have their own private willingness to sell (WTS), which determines the minimum price at which the supplier is ready to sell.  For supplier $i$, denote the WTS by $v^{(i)} = (v^{(i)}_1, ..., v^{(i)}_k)$. 
%: explain - like the volume discount bids?
%\st{The volume discounts are offered via the use of $k$ intervals, each with $\ell = \lfloor \frac{m}{k} \rfloor$ units. Thus, the intervals allow suppliers to specify the prices (per additional unit) for every interval, where the intervals are $[1, \ell], [\ell+1, 2\ell], ..., [(k-1)\ell + 1, k\ell]$ units respectively.} 
 These valuations are assumed to be drawn from some prior  distribution $\mathcal{F}$. In our setting,  $\mathcal{F}$ is common knowledge among all the suppliers and the buyer, whereas the realized vector of valuations $v^{(i)}$ is known privately only to the individual supplier $i$.   % The supplier $i$ then supplies bids $b^{(i)} = (b^{(i)}_1, ..., b^{(i)}_k)$. As long as the mechanism is incentive compatible, 
%The mechanism is incentive compatible if we  have $b^{(i)} = v^{(i)}$. %A supplier's \textbf{WTS} for some allocation, $a$ (where $a$ is the number of units bought from the supplier), is the minimum total price they would accept for that many units, i.e.
%\begin{gather}
 %   w (v, a) = \sum_{j=1}^{a} v_{\lfloor j/k \rfloor}
%\end{gather}
%where $v = (v_1, ..., v_k)$ is the private valuation tuple for that seller.

% Prathik: Looks OK to me^

The  utility for a supplier is defined as a function of her private valuations  $v^{(i)}$, allocation $a$, and payment $p$, and is given by
\begin{equation}
    u_i (v^{(i)}; b) = p_i(b) - \sum_{j=1}^{a_i(b)} v^{(i)}_{\lceil j/\ell \rceil}%.
\end{equation}
%\begin{gather}
%    u_i (v^{(i)}, a(b), p(b)) = p_i(b) - w(v^{(i)}, a(b))
%\end{gather}
% The function $w$ above is the the total valuation of the goods procured from agent $i$. We consider $w$ to be linear, i.e.,  
% Prathik: Linear in *what* precisely? It is a piecewise linear function in a_i(b)
% Also, this notation is somewhat unclear
% We could consider using entirely different letters here if necessary
% New notation if we don't use extra symbols:
% \begin{gather}
%    w (v, a) = \sum_{j=1}^{a} v_{\lfloor j/\ell \rfloor}.
% \end{gather}
% But this has its own problems. So with different symbols:
% \begin{gather}
% w (\nu, \alpha) = \sum_{j=1}^{\alpha} \nu_{\lceil j/\ell \rceil}.
% \end{gather}
% Here, $\nu = (\nu_1, ..., \nu_k)$ is the valuation tuple for that agent and $\alpha$ is the allocation for that agent.
%\begin{gather}
%    w (v^{(i)}, a(b)) = \sum_{j=1}^{a_i(b)} v_{\lfloor j/\ell \rfloor}.
%\end{gather}
%: Is ^i missing over v in RHS.

%\textcolor{red}{We train our model to be Dominant Strategy Incentive Compatible. (needed?) Expand if space available, remove otherwise} % This seems unnecessary to mention as we give more details regarding this in the part on regret. Yeah right, I am just worried about the continuity. It feels like the definition suddenly came.
We now recall some definitions. A mechanism is DSIC if no agent can gain utility by misrepresenting its valuations, regardless of the strategies adopted by the other agents. That is, 
\begin{equation}
    u_i (v^{(i)}; (v^{(i)}, b^{(-i)})) \geq u_i (v^{(i)}; b) \qquad \forall v, b, i. 
\end{equation}

% I think that line can be added
An auction mechanism is called ex-post IR if every agent earns non-negative utility by participating in the auction. We assume that there is no participation/entry cost.  Our proposed neural network architecture is designed to ensure the ex-post IR condition which is equivalent to 
%This is equivalent to the constraint:
\begin{equation}
    u_i(v^{(i)}; v) \geq 0 \qquad \forall v, i.
\end{equation}
%It is notable that ex-post individual rationality is the strongest form of individual rationality in mechanism design literature \cite{KRISHNA09}.

The goal is  to minimize the \textbf{cost} incurred by the buyer, subject to DSIC and ex-post IR for the sellers. The total cost to the buyer is given by
\begin{equation}
    \cost = \sum_{i=1}^n p_i(b).
\end{equation}
We remark here that the cost minimization under {DSIC} and {IR} constraints   corresponds to a nontrivial extension of the celebrated revenue optimal auction theory  \cite{myerson} (Myerson's auction only considers a single item and a weaker form of incentive compatibility, namely Bayesian incentive compatibility). Our approach considers additional,  practically motivated  constraints (multiple units, volume discount bids, envy minimization, and business constraints). 
Following section 2.2.2 of \cite{DUTTING21}, one can guarantee DSIC property by ensuring that the expected ex-post regret for every supplier, $r_i$, is $0$. The expected ex-post regret of a procurement mechanism is defined as 
\begin{equation}
        r_i = \mathbb{E}_{v \sim  \mathcal{F}}[\max_{b} [ u_i(v^{(i)}; b) - u_i(v^{(i)}; (v^{(i)}, b^{(-i)}))]].
\end{equation}
The regret is computed empirically, which adequately approximates the real regret \cite{DUTTING21}.
\newline \newline \noindent 
{\bf Remarks:}
\begin{itemize} [leftmargin={*}] %Comment out the leftmargin=* if space permits 
    \item The mechanism solicits volume discount bids from each supplier. 
    These bids represent each agent's  valuation for a single unit from each  `lot' of units. There are $k$ lots of (almost) equal size. The supplier $i$'s bid  $b_1^{(i)}$ is applied to all the goods if the buyer procures at most one lot; i.e. $\ell$ goods. For the  goods from the second lot,  the bid $b_2^{(i)} $ is used, i.e., a price of $b_1^{(i)}$ per unit for the first $\ell$ goods and a price of $b_2^{(i)}$ for the remaining goods (up to $\ell$). And so on.  %   If buyer procures $m_i$ lots  from the supplier the    in the first lot  
    \item All the goods from  a given lot are valued  equally. Thus, if the supplier sources more lots, her valuation per good decreases. In the agricultural domain, this corresponds to savings from the use of bulk transport, warehouse clearance, mass production, etc. 
    \item In many practical settings, the value of $k$ is determined endogenously. This value depends on factors such as packaging method, carton size, nature of the goods, etc. However, when $k$ is a design parameter, it presents  an interesting challenge in auction design. A larger value of $k$ introduces more granularity, which is better for the buyer. But it also introduces complex bidding procedures possibly leading to a less effective implementation. We leave the study of this aspect as an interesting future work. 
%: Mention here or after the example below, that when different suppliers have different k, we take the LCM. Yes. But time taken scales linearly in k.
\end{itemize}

\noindent
{\bf Example 2}: %Note that this simple model can be used to encapsulate a wide variety of situations. Following the previous example, 
Recall Example 1 where we had the supply curve: ((1-500: 20), (501-1500: 18), (1501-2500: 16)).  Here $m = 2500$. Suppose $k = 5$.  We will have $\ell=500$. The intervals would be $[1, 500]$, $[501, 1000]$, $[1001, 1500]$, $[1501, 2000]$, $[2001, 2500]$. 
Let the first supplier's WTS be $v^{(1)} = (20, 18, 18, 16, 16)$. If $1800$ items are  acquired from the first  supplier, the supplier's WTS for the allocation would be $500 \times 20 + 500 \times 18 + 500 \times 18 + 300 \times 16 = 32800$. Now assume that the second supplier has a production capacity of  $1500$  units, and she offers  a fixed volume discount after $500$ units. That is,  $v^{(2)} = (18, 17, 17, \infty, \infty)$. If the allocation for the second supplier is $700$ units, then her WTS for the allocation is $500 \times 18 + 200 \times 17 = 12400$. Thus, the suppliers may have a wide variety of different supply curves and still fit within the setting considered in this paper.

%\subsubsection{Comparison with Combinatorial Auctions}
%
%Volume discount setting has many desirable properties.  First, the volume discount setting considerably simplifies the bidding language of the auction \cite{boutilier2001bidding} and hence finds use in many practical applications including agri-inputs procurement. Second, as it is often the case that high volume procurement is regulated by government agencies, we show that the volume discount setting can easily accommodate these business constraints.

\subsection{Deep Learning Based Formulation}
We propose a neural network based formulation to obtain {DSIC} and {IR} guarantees along with some other desirable and practical constraints.  
The goal is to minimize a composite  loss function that consists of the  following three parts. The objective
cost (i.e., the total payment), the regret penalty, and the Lagrangian term (as we use the method of differential multipliers \cite{NIPS1987_CDO}) for regret. 
%\textcolor{red}{The Lagrangian parameter $\lambda_{{\tt regret}}$ is itself updated every few batches.} \textcolor{red}{We start with a small value of $\lambda_{{\tt regret}}$ so that the network initially focuses on the cost objective. Once the network has learned to minimize the cost, we update $\lambda_{{\tt regret}}$ to a larger value to ensure that the network projects the learned parameters to the required DSIC constrained region.} \textcolor{blue}{Ganesh: I am unsure what we mean by above sentense. Bazil,Prathik can you elaborate it a bit more? Answered above.} 
In particular, we have  %\newline 
\iffalse Thus:\\[1mm]
\begin{itemize}
    \item Objective function: $cost$

\item  Regret penalty which is  high  when the model deviates from IC constraint 
\item Lagrangian loss 
%\textit{Goal}: Minimize $cost$ subject to $r_i = 0, \forall i \in [n]$
\end{itemize}
\fi 
%MRB: \rho and \lambda used in box below but not defined
\newline 
\begin{center}
\fbox{ 
  \parbox{0.5 \linewidth}{
\begin{align*}
    { \tt loss} = { \tt cost } + { \tt penalty_{regret}} &+ { \tt LagrangianLoss}  \\
    { \tt cost} &= \sum_{i=1}^n p_i(b)\\
    { \tt LagrangianLoss} &= \sum_{i=1}^n \lambda_{{\tt regret}}^{(i)} \tilde{r}_i \\
    {\tt penalty_{regret}} &= \rho_{{\tt regret}} \sum_{i=1}^n \tilde{r}_i^2
\end{align*}
}}
\end{center}
%\newline \newline 
Here, $\tilde{r_i}$ is the empirical regret. We compute $\tilde{r_i}$ by using another optimizer over the bids, coming from the same distribution as $\mathcal{F}$, which maximizes the utility for agent $i$. To approximate the expectation over the distribution $\mathcal{F}$, we maximize the sample mean of regret over the batch. We discuss more on this subsequently.
%\textcolor{green}{We refer the reader to section 2.2.2 of \cite{DUTTING21} for more information.}

% \begin{figure*}[ht]
% \centering
% \begin{subfigure}
%   \centering
%   \vspace{3pt}
%   \includegraphics[width=0.49\linewidth]{AllocationNet.pdf}
% %  \caption{A subfigure}
% %  \label{fig:sub1}
% \end{subfigure}%
% \begin{subfigure}
%   \centering
%   \includegraphics[width=0.49\linewidth]{PaymentNet.pdf}
% %  \caption{A subfigure}
% %  \label{fig:sub2}
% \end{subfigure}
% \caption{Allocation and Payment networks}
% \label{fig:one}
% \Description{The Allocation network is displayed on the left while the figure on the right shows the Payment network}
% %    \centering
%  %   \includegraphics[width=0.49 \textwidth]{AllocationNet.pdf} 
%   %  \includegraphics[width=0.49 \textwidth]{PaymentNet.pdf} 
%    % \caption{Proposed RegretNet Architecture}
% \end{figure*}

\subsubsection{Business Constraints}
The buyer may wish to impose various business constraints in the procurement auction - for example, the buyer may require that at least 3 suppliers supply at least $20\%$ of the items each. We take care of this by adding a penalty term for violating various business constraints while training the network. For having a minimum of $s$ suppliers each with an allocation of at least $a_{min}$, the penalty would be
\begin{equation}
    { \tt penalty_{business}} = \begin{cases}
        0 & a^{(s)} \geq a_{min} \\
        \frac{\rho_{{\tt business}}}{a^{(s)}} & a^{(s)} < a_{min}
    \end{cases}
\end{equation}
where $a^{(s)}$ is the $s^{th}$-highest allocation. 
\iffalse 
\begin{align*}
    { \tt loss }   = {\tt cost}   + {\tt   penalty_{regret}} \  + \  { \tt penalty_{business} } \\
    \qquad { \tt + \  LagrangianLoss.}
\end{align*}
\fi
Other business constraints are also possible. For instance,   no supplier is allocated more than $50\%$ of the units. The loss function in this case also adds the penalty for violating  business constraints.

%A possible improvement exists by adding a penalty for all suppliers among the $s$-highest. This improvement could be mentioned in Further Work (but is an implementation detail). Other business constraints can be mentioned there as well.

\subsubsection{Envy Minimization}
In auctions, it is often desirable to have some additional fairness constraints, specifically, minimization of envy. Envy (or dissatisfaction) for an agent is defined as the maximum utility they could gain if they were given the allocation and payment of some other agent. So the envy for supplier $i$, given the valuation tuple $v = (v^{(1)}, ..., v^{(n)})$ is
% for space savings, I use u_i instead of writing the full thing
\begin{align}
        e_i (v) = \max_{h\in[n]}& [(p_h(b) - \sum_{j=1}^{a_h(b)} v^{(i)}_{\lceil j/\ell \rceil})] - u_i (v^{(i)}; v)
\end{align}
We minimize envy by adding a term for envy in our Lagrangian loss, along with an envy penalty.   \newline \newline 
\begin{center}
\fbox{
  \parbox{0.65 \linewidth}{
\begin{align*}
  \loss = \cost + { \tt penalty_{regret}} &+ { \tt penalty_{envy} } + { \tt LagrangianLoss}\\
    { \tt cost} &= \sum_{i=1}^n p_i(b)\\
    { \tt LagrangianLoss } &= \sum_{i=1}^n 
    \lambda_{{ \tt regret}}^{(i)} \tilde{r}_i + \lambda_{{\tt envy}}^{(i)} e_i \\
    { \tt penalty_{regret}} &= \rho_{{ \tt regret}} \sum_{i=1}^n \tilde{r}_i^2 \\ { \tt penalty_{envy}} &= \rho_{{\tt envy}} \sum_{i=1}^n e_i^2
\end{align*}
}}
\end{center}

\subsubsection{Business Constraints With Envy Minimization}
If we wish to have business constraints while simultaneously minimizing envy, the loss function for the network is %\newline \newline 
\begin{center}
\fbox{
  \parbox{0.9 \linewidth}{
  \begin{align*}
   {\tt loss \ = \ cost \ }  {\tt + \  penalty_{regret} \ } & { \tt \ + \   penalty_{envy} } + { \tt penalty_{business}} + { \tt LagrangianLoss} \\
     { \tt cost} &= \sum_{i=1}^n p_i(b)\\
    { \tt LagrangianLoss} &= \sum_{i=1}^n  \lambda_{{ \tt regret}}^{(i)} \tilde{r}_i + \lambda_{{ \tt envy}}^{(i)} e_i \\
    { \tt penalty_{regret}} &= \rho_{{ \tt regret}} \sum_{i=1}^n \tilde{r}_i^2 \\ { \tt penalty_{envy}} &= \rho_{{ \tt envy}} \sum_{i=1}^n e_i^2 \\
    { \tt penalty_{business}} &= \begin{cases}
        0 & a^{(s)} \geq a_{min} \\
        \frac{\rho_{{\tt business}}}{a^{(s)}} & a^{(s)} < a_{min}
    \end{cases}
\end{align*}
}}
\end{center}

\subsection{Allocation Network and Payment Network}\label{subsec:NetDetails}
\begin{figure}[ht]
        \centering
        %\framebox{\parbox{3in}{\includegraphics[scale=0.36]{AllocationNet.pdf}}}
        \includegraphics[scale=0.65]{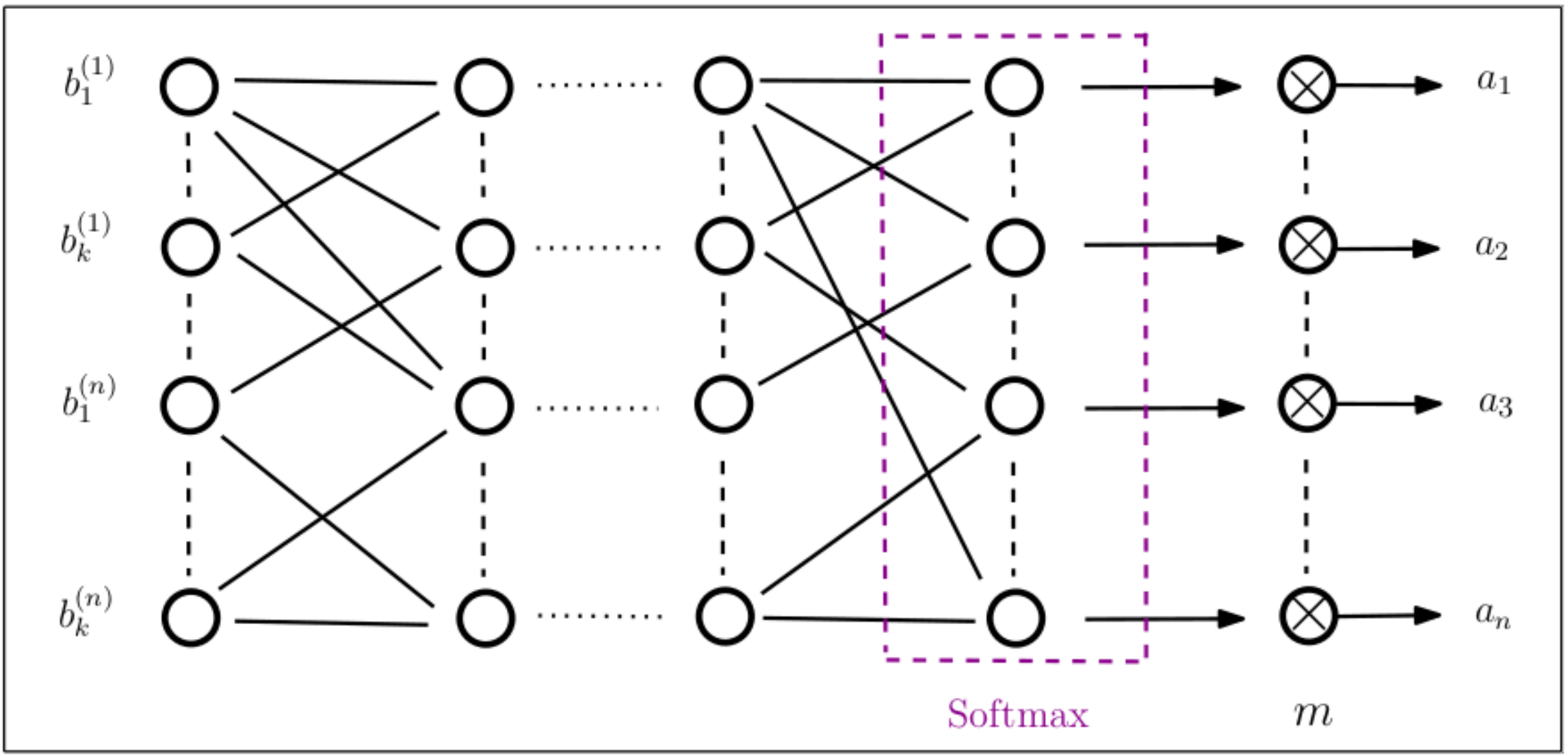}
        \caption{Allocation Network}
        \label{fig:allocationNetwork}
    \end{figure}

\begin{figure}[ht]
        \centering
        %\framebox{\parbox{3in}{ \includegraphics[scale=0.36]{PaymentNet.pdf}
        %}}
        \includegraphics[scale=0.65]{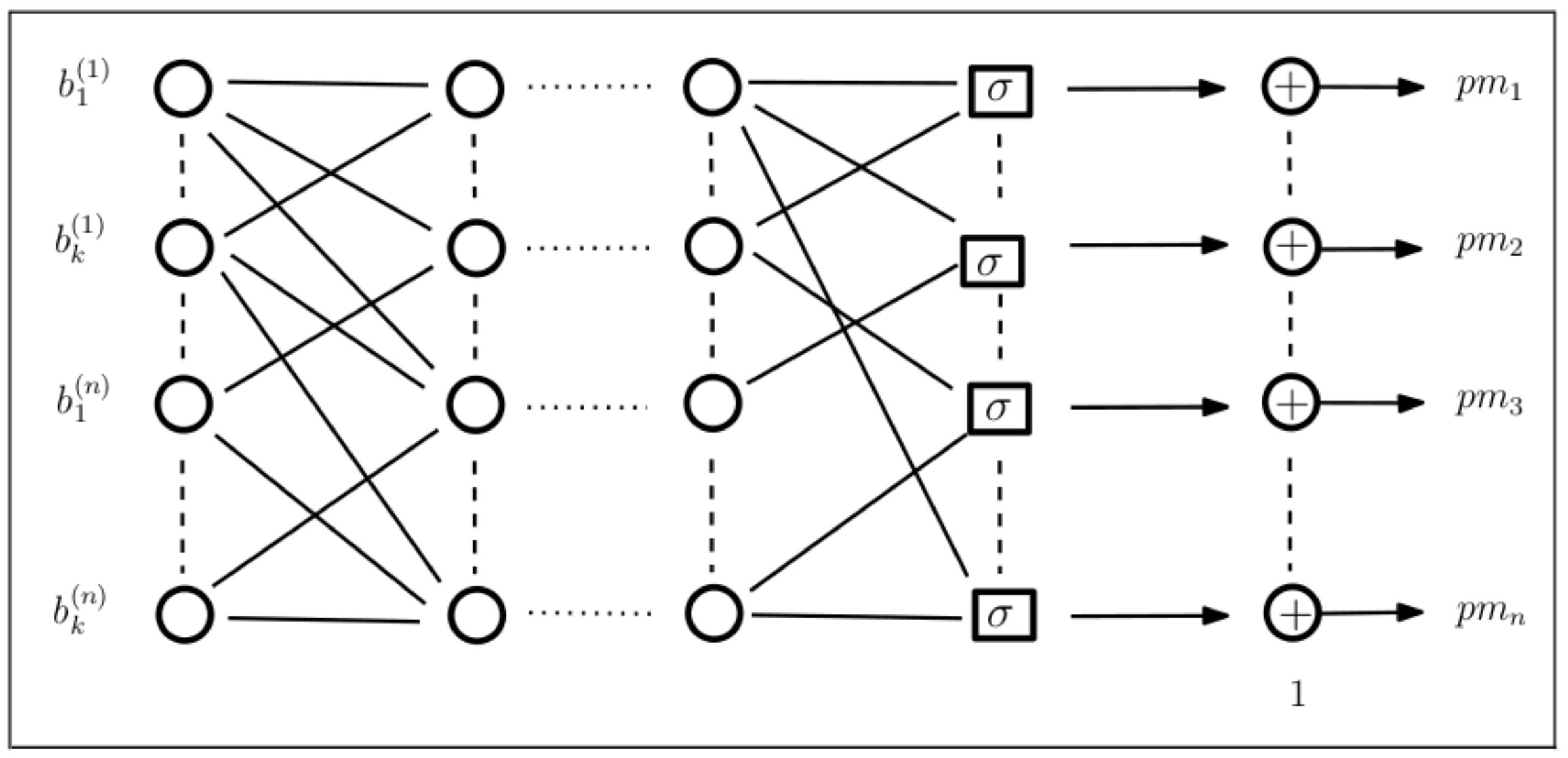}
        \caption{Payment Network}
        \label{fig:paymentNetwork}
    \end{figure}

The model consists of two feed-forward networks - an allocation network and a payment network (See Fig. ~\ref{fig:allocationNetwork}, Fig. \ref{fig:paymentNetwork} for details).  The input for both networks is the $n \times k$ matrix where the $i^{th}$ row is the bid $b^{(i)}$ for supplier $i$, which is assumed to be equal to the valuation. The output of the allocation network is the allocation tuple described in Section \ref{subsec:Setting}. The allocation network uses soft-max to ensure that the allocation tuple is a probability vector. This is multiplied by $m$ to ensure that the allocations across the agents sum up to exactly $m$. The output of the payment network is a payment multiplier tuple, $\hat{p} = (\hat{p}_1, ..., \hat{p}_n)$, %(\textcolor{red}{This notation 'pm' is little confusing, as we already have p and m in our notion, representing payment and units to be procured. So instead of using 'pm' for payment multiplier we can use single character notation like it was there in dutting et al, $\hat{p}$})
which, when multiplied by the total WTS of the allocation, gives the payment tuple, i.e.,
\begin{equation}
    p_i (b) = \hat{p}_i (b)\sum_{j=1}^{a_i(b)} v^{(i)}_{\lceil j/\ell \rceil}
\end{equation}
Each $\hat{p}_i$ is guaranteed to be within the range $[1, \infty)$ in order to ensure IR. The network architecture ensures this by using a sigmoid layer followed by the addition of the constant $1$, thus constraining the output to $(1, 2)$.

%\textcolor{green}{In all our experiments, $4$ or $5$ layers, with $60$ to $150$ neurons in each layer, were used  for both the payment and allocation networks. The Adam optimizer was used for training the network weights, while stochastic gradient descent was used to learn the Lagrangian parameters.}

% Should we explain this? We can mention that the network architecture ensures this. looks good but, above should we explain why are we restricted to (1,2)instead of (1,infinity)?
% Alright, I'll do that then
%I guess we should
%\textcolor{red}{We cannot use the RegretNet architecture off-the-shelf since 1) The payment rule does not include volume discount since  payment for each item  is computed independent of other items 2) They use the fact that valuations are additive 3) Ours is a procurement auction so the requirement of individual rationality is different   }

%\textcolor{red}{1) Their allocation is a matrix. For each item they give a distribution. Hence it is a $n\times m$ matrix. As we have homogeneous items we allocate fraction of items given by the distribution... 2) The way payment is computed is different, (need to add this to the figure) 3) To maintain individual rationality they use the range of payment multipliers is set as  [0,1] whereas this doesnt work in our case since we are using multiple items. The payment multiplier needs to be a larger value (we use the range [1,2]) 4) We use ReLu over sigmoid in the last layer to obtain the required feasible range for  payment multipliers.     }
%Bazil_New_Change

\subsection{Training Procedure}
In all our experiments, $4$ or $5$ layers, with $60$ to $150$ neurons in each layer, were used  for both the payment and allocation networks. The Adam optimizer was used for training the network weights, while stochastic gradient descent was used to learn the Lagrangian parameters and compute the empirical regret.

During the training phase, we performed nested optimizations. For one step of optimization over the network weights, we executed  $R$ steps of optimization over the bids, to compute the empirical regret. We did gradient ascent over the Lagrangian parameters after every 2-4 epochs with gradual increase in learning rate over the epochs. This was done to ensure that, first, the network weights converge towards minimizing the cost objective and then, the learned weights get projected in the constrained region to satisfy other required properties of the auction. The above idea for empirical regret computation 
is the same as the one proposed in \cite{DUTTING21}.
%has been directly borrowed from \cite{DUTTING21}. 
%We refer the reader to section 2.2.2 of \cite{DUTTING21} for more information on this.
\subsection{Some Notes on the Methodology}
\noindent
{\bf Versatility}: The methodology presented in this paper is versatile in the sense of its ability to model the minimization or maximization of a wide variety of performance metrics. For example, we can use this methodology to  maximize the Nash social welfare subject to incentive compatibility, individual rationality, and business constraints. Nash social welfare maximization is widely known for its fairness properties \cite{caragiannis2019unreasonable}. The techniques presented in \cite{HOHNER03, BICHLER05, CHANDRASHEKAR07, IYENGAR08, GAUTAM09} correspond to  very specific settings and are not generalizable. The approach presented here offers many powerful features that can be modeled in volume discount auctions.
\newline\newline
\noindent
{\bf Computational Complexity}: The methodology proposed here essentially transforms a mechanism design problem into an optimization problem. So the question would arise as to why an efficient optimization procedure cannot be used to solve the problem, at least approximately. The constraints we deal with such as business constraints and envy minimization are nonlinear and linear approximations do not work well with such constraints. Moreover, the linear approximation will have an exponential number of variables. The deep learning technique has the advantage that a single substantial effort of training will amortize the computational complexity over a large number of experiments. 
%Researchers have earlier explored automated mechanism design (AMD). The basic idea of AMD is to computationally search through the space of feasible mechanisms rather than designing it analytically by hand \cite{10.1145/2559049}. Traditionally, AMD has been handled using \textbf{Linear Programming (LP)} [For example, \cite{conitzer2002complexity,conitzer2004self}]. The LP technique is not  scalable because of the exponential growth in the number of variables and constraints in the  number of bidders and items \cite{guo2010computationally}. One way to make AMD more tractable is to search in a parameterized subfamily of feasible mechanisms \cite{conitzer2002complexity}. Neural networks, being a universal function approximator, have shown promising results in parametrized mechanism design \cite{DUTTING21,manisha2019thompson}.

%%%%%% Section 3
\section{Experimental Results}%for Mechanisms Designed}
\label{sec:ExperimentalResults}
In this section, we present experimental results for the following six auctions. Notably,  all these auctions satisfy DSIC and IR. 
\begin{enumerate}
    \item A standard VCG auction (satisfies SWM)
    %(this is strategy proof and individually rational and maximizes social welfare)
    \item A VCG auction subject to a limit on the minimum number of winning suppliers (satisfies SWM and BUS)
    \item A cost minimizing volume discount auction (satisfies OPT)
    \item A cost minimizing volume discount auction with envy minimizing allocation (satisfies OPT and FAIR)
    \item A cost minimizing volume discount auction with a limit on the minimum number of winning suppliers (satisfies OPT and BUS)
    \item A cost minimizing volume discount auction with envy minimizing allocation and with a limit on the minimum number of suppliers (satisfies OPT, FAIR, and BUS)
\end{enumerate}

In this experiment, we have $5$ suppliers.  $5$ is a realistic number in agri-settings.  In a high volume agri-input procurement, the number of qualified suppliers is typically small. %It is unusual to have a large number of qualified suppliers for bulk procurement. 
Also, an initial qualification process can be used to weed out suppliers who do not satisfy the quality requirement. This is a definite advantage of bulk procurement; since the FC negotiates the quality and cost, a desired quality and a competitive price are assured. Left to themselves, individual farmers have no bargaining power on either quality or price.%, due to the intermediaries. 

In the current experiment, we assume $5000$ or $10000$ units  procured from the suppliers who submit volume discount bids. Each bid will have a set of intervals and a certain discount corresponding to that interval. We use a minimum interval size of $500$ and each interval in a bid has a size that is an integer multiple of the minimum interval size. We assume a base price of US \$ $5$ per unit, and a minimum profit margin that ranges from $10\%$ to $30\%$. Recall that the base price plus the minimum profit margin is the willingness to sell of the supplier. The volume discounts typically range from $1\%$ at the first discounted interval to as high as $25\%$ on the highest discounted interval. The data were synthetically generated based on our experience with the field visits we had undertaken to two FCs.  For business constraints, we assume at least $3$ winning suppliers have to be selected for procurement. This helps introduce redundancy for decreasing fragility in the supply chain. 

Table \ref{table:1} shows the total procurement cost with six different auctions.  We ran the simulation for over 12000 instances, randomly generated from the same distribution as of the training data, and averaged the results to populate the table. We dynamically generate these data to ensure that the model has never seen these data during training. This shows that our model generalizes over the data distribution rather than just over-fitting a particular set of instances.

%The results given are computed as  averages over 200 runs with bid amounts generated according to a uniform distribution around the respective base values for different parameters. We ran 200 simulation experiments with random numbers generated for all appropriate variables and took an average over those 200 simulations.

%This one was already in USD
% Rather, it had no units attached so we can just say USD
\begin{table}[h]
\centering
%\begin{tabular}{c p{4cm} c c}
\begin{tabular}{c l c c}
    \hline %\toprule
    & {\bf Auction Type} &\textbf{5000 Units}  & \textbf{10000 Units} \\
    \hline %\midrule
    1 & VCG Auction & $27935$ & $52930$ \\
    \hline
    2 & VCG Auction + Business Constraints & $30550$ & $60100$ \\
    \hline
    3 & Cost Minimizing & $27550$ & $51950$ \\
    \hline
    4 & Cost Minimizing +  Envy Minimizing & $27700$ & $52350$ \\
    \hline
    5 & Cost Minimizing + 
    Business Constraints & $29050$ & $56800$ \\
    \hline
    6 & Cost Minimizing +  
    Envy Minimizing + 
    Business Constraints & $29600$ & $54050$ \\
    \hline \\ %\bottomrule  
\end{tabular}
\caption{Total procurement cost in US \$}
\label{table:1}
\end{table}

Clearly, Auction (6) is the most desirable. It has a cost higher than that of a VCG auction but lower than that of a VCG auction with a limit on the number of winning suppliers.  Auction (6) clearly will have a cost lower than that of a VCG auction with envy minimization and limited number of winning suppliers. As expected, Auction (3) (cost minimizing) has the least cost.   
%It is important to minimize the total cost of procurement since it directly benefits the individual farmers who are hard pressed for money. 
If envy minimization and business constraints are not a consideration, we go for Auction (3). If envy minimization is important but not business constraints, we go for Auction (4). If business constraints are important but envy minimization can be ignored, we go for Auction (5). If all properties are important, we go for Auction (6). The deep learning based methodology thus enables different options to be exercised based on the context. The key direct benefit is reduction of cost to farmers and the key indirect benefit is assurance of minimum quality.

%%%%%%%%%%%%%%%%% Section 4
\section{Two Case Studies: Chili Pepper Seeds and a Popular Pesticide}\label{sec:CaseStudies}
%\section{TWO CASE STUDIES: CHILI PEPPER SEEDS AND A POPULAR PESTICIDE}\label{sec:CaseStudies}
%In order to gain first-hand experience and knowledge of the real situation on the ground, our research group undertook a field study of two FCs, approximately 1 hour drive from the university campus, in different directions.  Both these FCs deal with horticulture and have a membership of about 1000 farmers each and  are  proactively engaged in helping out farmers in various ways. 
%: This is a clear give-away of who we are, considering that the COMPASS paper mentions both the FCs by name
In a typical farmer collective the farmers usually approach intermediaries for sourcing their inputs.  The intermediaries allure the farmers by providing credit for sourcing the agri-inputs. In the process, the intermediaries  are able to capture the  marketing and selling of the produce as well, with huge commissions, causing severe loss of revenue to the farmers.  Small and marginal farmers tend to be low on education and are particularly vulnerable to the selfish moves of the intermediaries. This is where the FCs can help; the FCs can play a key role in streamlining the supply of inputs to the farmers and counter the intermediaries. 
%When we visited the two FCs, which have done excellent work, we saw for ourselves the level of confidence that farmers had in the two FCs and the eagerness of new farmers to enroll themselves as members of the FCs. 

%Our team had  a discussion with the FCs on how they aggregate the input requirements of the farmers and bulk-procure the right quantities of inputs to be sold subsequently to the farmers at affordable prices. 
%During these conversations, we also realized numerous issues which were hampering a successful execution of the bulk procurement process. For example, FCs do need a healthy amount of money and resources to execute this process more efficiently. 
%We also discussed with the FCs the discounts that they would be able to obtain because of the volume of their purchases. Here again, we found that the discounts on offer from the suppliers could be much higher if the right kind of procurement protocols are put in place. The mechanisms developed in this paper provide an attractive option.
%With this as the motivation, our paper now explores procurement auctions with volume discount bids as a competitive mechanism that FCs can deploy.

\subsection{Procurement of Chili Pepper Seeds}
%: This para Gives away that we are from IISc
Our first case study is on chili pepper. This is inspired by the study presented in \cite{BHARDWAJ22}. 
%Chili pepper is the dried ripe fruit of the genus capsicum (Capsicum annum L). It is an annual subshrub and an important commercial spice crop in any region with  warm humid climate. We repeat some essential details of this case study from \cite{BHARDWAJ22}.
There are numerous varieties of chili pepper seeds (more than 50).
%These could be grouped under five main varieties. The FC can use a mobile app to collect the individual requirements of the farmers. 
We consider here two types of seeds (A and B). The seeds come in packets of 4 kg each. There is a demand of 2000 packets for seeds of type A while there is a demand of 1000 packets for seeds of type B.  Call each packet a unit. 
The FC can  bulk-procure this requirement from major suppliers of these seeds and then distribute the required volume of seeds to farmers at affordable prices. 

We consider a volume discount auction where each volume discount bid has four equal segments offering 
%2.5 percent, 5 percent, 7.5 percent, and 10 percent 
2.5\%, 5\%, 7.5\%, and 10\%
discount on per unit price. For example, in the case of seeds of type A, the 2000 units are divided into four segments, namely, [1,500], [501,1000], [1001, 1500], and [1501, 2000] and in these segments, the discounts offered are 2.5\%, 5\%, 7.5\%, and 10\%, respectively of the bid amount when discounts are not offered.

Table \ref{table:2} presents the results for all six types of auctions separately for type A seeds and type B seeds.  These results are computed as  averages over 12800 samples with bid amounts 
drawn from 
%generated according to 
a uniform distribution around the respective base values for different parameters. These results are computed using a trained model, but on data generated separately from the data used for training the model. In all cases, the interval size, $l$, was fixed to be $100$. The base prices for \textbf{A} and \textbf{B} are \textbf{\$17.11} and \textbf{\$14.47} respectively. The minimum profit margin (in \%) was assumed to be distributed uniformly over $[8, 12]$. Here, we assume there are \textbf{10} suppliers.

\begin{table}[h]
\centering
%\begin{tabular}{|c|p{4cm}|c|c|}
%\begin{tabular}{c p{5cm} c c}
\begin{tabular}{c l c c}
    \hline %\toprule
    & \textbf{Auction Type} & \textbf{2000 A} & \textbf{1000 B} \\
    \hline %\midrule
    1 & VCG Auction & $34880$ & $14750$ \\
    \hline
     2 & VCG Auction +  
    Business Constraints & $36000$ & $15220$ \\
    \hline
        3 & Cost Minimizing & $34760$ & $14700$ \\
    \hline
    4 & Cost Minimizing + 
    Envy Minimizing & $34960$ & $14775$ \\
    \hline
    5 & Cost Minimizing + 
    Business Constraints & $34770$ & $14710$ \\
    \hline
    6 & Cost Minimizing +  
    Envy Minimizing + 
    Business Constraints & $35225$ & $14960$ \\
    \hline \\%\bottomrule
\end{tabular}
\caption{Chili Pepper seeds: Total procurement cost in US \$}
\label{table:2}
\end{table}

%From the table, 
From Table \ref{table:2}, it is clear that Auction (6) is the most desirable. 
%It has a cost higher than that of a VCG auction but lower than that of a VCG auction with a limit on the number of winning suppliers.  Auction (6) will obviously have a cost lower than that of a VCG auction with envy minimization and limited number of winning suppliers. As expected, Auction (3) (cost minimizing) has the least cost.   
%It is important to minimize the total cost of procurement since it directly benefits the individual farmers who are hard pressed for money. 
In the current context of chili pepper seeds,  envy minimization and business constraints are both important and we go for Auction (6). Note that this is an auction that (nearly optimally) satisfies dominant strategy incentive compatibility, ex-post individual rationality, envy minimization, business constraint of minimum number of winning suppliers, and cost minimization.
The deep learning based methodology thus enables the best option to be exercised, with the direct benefit of reduction of cost to farmers and the indirect benefit of assured quality of chili pepper seeds.

\subsection{Pesticide Procurement}
%As already stated in Section \ref{sec:Intro}, the use of pesticides in the right quantities at the right time saves crops from being wiped away. Pesticides constitute a substantial share of agri-inputs bought from the FCs by the farmers. There are numerous types of pesticides and 
For our experimentation, we have chosen a popular pesticide.%, which we prefer to keep anonymous. 
This pesticide  comes in packets of 250 grams. A typical small farmer may need several packets of pesticides (say 5 to 10).  We assume 1000 farmers in the FC and a requirement of 5000 packets to be sourced from 5 suppliers who offer volume discounts. 
% (if space permits?)
%This is the ideal price to sell since it would fetch the cost (base price) plus desirable profit margin. Suppose the cost is Rs 250 and 20% is the desirable profit margin. Then Rs 300 is the WTS. This is the private information of the supplier. This will fix the maximum volume discount that the seller may offer in order not to make a loss.
%Suppose q is the desirable profit margin.
%For each supplier, we have cost (base price), capacity (max number of units that the seller can supply), desirable profit margin, and supply curve.
Let $S1$, $S2$, $S3$, $S4$, and $S5$ be the five suppliers. The following are the supply curves of these suppliers.
%
%You may please take the details as:
%Supplier   Cost   q       WTS         capacity 
%S1              200    10    220         3000 
%S2              210    15    241.5      3000 
%S3              220     20   264         4000 
%S4              230     30   299        4000 
%S5              200     50   240         5000
{%\small
\begin{verbatim}
S1: [1-500; 2.75];[501-1000; 2.69];[1001-2000; 2.62];[2001-3000; 2.56] 
S2: [1-100; 3.02];[101-300; 2.94];[301-700; 2.87];[701-1000; 2.81];  
    [1001-1400: 2.77];[1401-1800; 2.75];[1801-2200: 2.72];[2201-3000; 2.69] 
S3: [1-500: 3.30];[501-1500; 2.87];[1501-3000: 2.81];[3001-4000; 2.77] 
S4: [1-500: 3.74];[501-1000; 3.25];[1001-2000; 3.00];[2001-4000; 2.87] 
S5: [1-5000; 3.00]  
\end{verbatim}}
%Note that different suppliers have different capacities and different volume discount offerings. 
These supply curves have been formulated after conversations with a few pesticide suppliers. The results given are computed using a trained model specifically tuned for this task.

We also provide some additional information for Auction (3). The same for the other auctions is omitted for the sake of brevity. The model provided an allocation of $[2937, 1980, 83, 0, 0]$ and a payment of $[8299, 5731, 274, 0, 0]$ for the 5 suppliers (in order). The total payment is thus $14304$ (rounded to $14300$). This is in contrast to Auction (1) where the allocation is $[3000, 2000, 0, 0, 0]$ and the payment is $[8626, 5931, 0, 0, 0]$ for the 5 suppliers (in order). Thus, total payment for VCG is $14557$ (rounded to $14560$).  

% P/B: Let us add this statement(consider rephrasing):
% \textcolor{blue}{PS: In this particular case, the effects of business constraints are negligible because of the relative similarities between the supply curves of S2, S3 and S5 at the quantities being considered, combined with the fact that S1 is simply the best overall option, but cannot supply 5000 units. In general adding business constraints makes a significant difference.}

Table \ref{table:3} presents the results for all six types of auctions. 
The trends exhibited are identical to the case of chili pepper seeds. Here again, envy minimization and business constraints are both important and we go for Auction (6). The deep learning based methodology thus enables the best option to be exercised, with the direct benefit of reduction of cost to farmers and the indirect benefit of assured quality of pesticide procured. 

\begin{table}[ht]
\centering
%\begin{tabular}{|c|p{5cm}|c|}
\begin{tabular}{c l c}
    \hline %\toprule
    & \textbf{Auction Type} & \textbf{Cost (in US \$)} \\
    \hline %\midrule
     1 &  VCG Auction & $14560$ \\
    \hline
    2 &  VCG Auction  + 
    Business Constraints & $14710$ \\
    \hline
    3 & Cost-Minimizing & $14300$ \\
    \hline
    4 & Cost-Minimizing + 
    Envy-Minimization & $14490$ \\
    \hline
    5 & Cost-Minimizing + 
    Business Constraints & $14390$ \\
    \hline
    6 & Cost-Minimizing + 
    Envy-Minimization  + 
    Business Constraints & $14640$ \\
    \hline \\%\bottomrule 
\end{tabular}
\caption{Pesticide: Total procurement cost in US \$}
\label{table:3}
\end{table}

\section{Conclusions and Future Work}\label{sec:Conclusion}
In this paper, we have designed a powerful mechanism for procurement of agri-inputs by farmer collectives using volume discount auctions. 
%This demonstrates an attractive opportunity to help reduce input costs at assured quality levels for smallholder farmers through farmer collectives.
The designed auctions minimize the total cost of procurement subject to fairness constraints and business constraints.
Simulation experimentation on these auctions on synthetic data as well as two stylized case studies show the efficacy of the mechanisms designed. 
%The mechanisms are intuitive and  are easy to explain to the suppliers. Suppliers also relate to these mechanisms since they practice volume discounts  routinely in their transactions. A big plus point is that the proposed mechanisms guarantee certain nice properties and induce honest bidding behavior from the suppliers. 
%There is, however, a wide gap between a pilot study like this one and an actual practical implementation and demonstration. 
Our work provides clear evidence that the proposed mechanisms will be more cost-effective than existing traditional methods, in addition to many other benefits they bring in, such as ensuring quality of agri-inputs, inducing honesty in bidding, bargaining power, selecting deserving suppliers, and the possibility to ensure fairness of allocation.  

It is important to see how such mechanisms can be deployed. There is  euphoria about these mechanisms in the two farmer collectives that we have surveyed.   We are currently implementing and deploying these  auctions in the two FCs. Deploying these on the ground does pose a few challenges. A key challenge is to convince the farmer collective and the farmers that these mechanisms will indeed work. This is directly connected to the explainability of these mechanisms. We are currently working on this.

%In our ongoing and future work, 
%The simulation experiment itself was on fairly standard methods in procurement auctions. 
%There are many avenues for extending our work in various directions. In the bidding methods, one can generalize the structure of the combinatorial bids. We have assumed that each supplier places only one combinatorial bid; this is quite restrictive. There is rich literature on bidding languages in combinatorial auctions and this literature can be invoked for implementing more powerful auctions \cite{PALACIOS21}. On the payment side, %the VCG payments ensure dominant strategy incentive compatibility and maximize social welfare (sum of values of all the agents). However, 
%a VCG auction maximizes social welfare but may not minimize the total cost of procurement. In this connection, optimal volume discount auctions and optimal combinatorial auctions need to be looked into \cite{KRISHNA09} and this is a promising research direction.

%%%%%%%%%%%%%%%%%%%%%%%%%%%%%%%%%%%%%%%%%%%%%%%%%%%%%%%%%%%%%%%%%%%%%%%%%%%%%%%%

\section*{Acknowledgments}
The first author would like to thank the Government of India, Ministry of Education, for providing the doctoral fellowship. All the authors would like to thank the national Bank for Agriculture and Rural Development (NABARD), Government of India, for supporting this work. The fourth Author would like to thank the support from SERB grant CRG/2022/007927 for the support.

%Bibliography
\bibliographystyle{unsrt}  
\bibliography{AGRIBIB} 
%\bibliographystyle{IEEEtran}
%\bibliography{IEEEabrv,AGRIBIB}

\end{document}